\documentclass[runningheads]{llncs}
\usepackage[T1]{fontenc}
\usepackage{graphicx}
\usepackage{subcaption}
\usepackage{caption}
\usepackage{listings}
\usepackage{xcolor}
\usepackage{url}
\usepackage{svg}
\usepackage{algorithm}
\usepackage{algpseudocode}
\usepackage{wrapfig}

\definecolor{commentsColor}{rgb}{0.497495, 0.497587, 0.497464}
\definecolor{keywordsColor}{rgb}{0.000000, 0.000000, 0.635294}
\definecolor{stringColor}{rgb}{0.558215, 0.000000, 0.135316}

\DeclareMathAlphabet{\pazocal}{OMS}{zplm}{m}{n}
\newcommand{\unif}{\pazocal{U}}

\begin{document}
\title{Programming abstractions for preemptive scheduling in FPGAs using partial reconfiguration}
\author{Gabriel Rodriguez-Canal\inst{1} \and
Nick Brown\inst{1} \and
Yuri Torres\inst{2} \and
Arturo Gonzalez-Escribano\inst{2}}
\authorrunning{Rodriguez-Canal et al.}
\institute{EPCC at The University of Edinburgh, EH8 9BT Edinburgh, UK \\
\email{gabriel.rodcanal@ed.ac.uk} \\ %\email{n.brown@epcc.ac.uk} 
\and
Escuela de Ingeniería Informática at Universidad de Valladolid, 47001 Valladolid, Spain %\\
%\email{\{yuri.torres,arturo\}@infor.uva.es}
}

%\institute{Princeton University, Princeton NJ 08544, USA \and
%Springer Heidelberg, Tiergartenstr. 17, 69121 Heidelberg, Germany
\maketitle              % typeset the header of the contribution
\begin{abstract}
%The reconfigurability of FPGAs provide the possibility to create tailored hardware for HPC kernels that deliver high performance. However, the numerous challenges associated to their use has prevented FPGAs from being adopted by any of the large supercomputers. In this work, we address the challenge of flexibility, given by the lack of mechanisms for preemption and the need to stall the FPGA when a new kernel has to be loaded in run-time. We present a user-facing interface that unlocks the preemption of kernels whilst abstracting the low-level details of Dynamic Partial Reconfiguration and a transparent method for the generation of the necessary hardware. Our approach incurs only in 1.66\% overhead in one RR and 4.04\% in two RRs, whilst presenting a significant performance improve over the traditional full reconfiguration approach.

FPGAs are an attractive type of accelerator for
all-purpose HPC computing systems due to
the possibility of deploying tailored hardware on demand.
However, the common tools for programming and operating
FPGAs are still complex to use,
specially in scenarios where diverse types of tasks
should be dynamically executed.
In this work we present a programming abstraction
with a simple interface that internally 
%hidding the complexty of the low-level details.
leverages High-Level
Synthesis, Dynamic
Partial Reconfiguration and synchronisation mechanisms
to use an FPGA as a multi-tasking server with preemptive
scheduling and priority queues. This leads to a better
use of the FPGA resources, allowing the execution of
several kernels at the same time and deploying 
the most urgent ones as fast as possible.
The results of our experimental study show that
our approach incurs only a 1.66\% overhead 
when using only one Reconfigurable Region (RR), and 4.04\%
when using two RRs, whilst presenting a significant 
performance improvement over the traditional 
non-preemptive full reconfiguration approach.

\end{abstract}
\section{Introduction}

The end of Moore's law %\cite{moore1998cramming} 
and loss of Dennard's scaling %\cite{dennard2007design} 
has motivated the search of alternative ways of improving the performance of upcoming computational systems. As a result, heterogeneous systems, primarily composed of CPUs and GPUs \cite{top500}, have become commonplace in HPC machines. However, these architectures are not ideally suited for all codes, and it has been found that when HPC applications are bound by aspects other than compute, for instance memory bound codes, moving to a dataflow style and exploiting the specialisation of FPGAs can be beneficial \cite{brown2019s,brown2020exploring}. Nonetheless, FPGAs have not yet been adopted by any of the large supercomputers, which is due to both the challenges of programmability and flexibility. The former has been partially addressed by High Level Synthesis (HLS) tooling, enabling the programmer to write their code in C or C++. However the latter has been less explored. The entire FPGA is often stalled during fabric reconfiguration which means that dynamic scheduling and preemptive execution of workloads is less common. 

In this paper we propose a programming abstraction to easily use an FPGA
as a multi-tasking server with preemptive scheduling and priority queues.
It hides the complex low-level details of using Dynamic Partial Reconfiguration (DPR) and synchronisation mechanisms to support on-the-fly instantiation, stopping and resumming of kernels on parts of the FPGA fabric whilst the rest of the chip continues executing other workloads independently. The proposal includes, as a case study,
the development of a full First-Come-First-Served (FCFS) preemptive scheduler with priority queues.
The tasks are programmed as OpenCL kernels managed with the Controller model \cite{moreton2018controllers,rodriguez2021efficient}, a heterogeneous programming model implemented as a C99 library of fucntions. It is oriented to efficiently manage different types of devices with a portable interface. The Controller model has been extended to support multiple kernels and preemption on DPR capable FPGA systems.
 %Finally, we provide a framework for the generation of such DPR system and the corresponding partial bitstreams for all the permutations of the kernels provided by the user.
This solution brings all the benefits of task-based models to FPGAs, with a low programming effort.
We also introduce an experimental study to show the efficiency of the proposed solution.

The rest of the paper is organised as follows: Section \ref{sect:rel_work} describes related activities tackling flexible execution of kernels for FPGAs.
Section \ref{sect:controller} presents an overview of the original programming model that we use as a base to
devise and implement our proposal. 
In Section~\ref{sec:approach} we present the techniques and extensions
to support our approach, both on the management of the on-chip FPGA infrastructure and on the host code.
Section \ref{sec:abstractions} describes the programming level abstractions provided to the user. In Section \ref{sec:experiments} we present an
experimental study to evaluate our approach. Section \ref{sect:conclusions} concludes the paper and discusses further work.

\section{Related work} 
\label{sect:rel_work}

The integration of different types of architectures in heterogeneous systems can enable the execution of workloads more efficiently by using the most appropriate hardware for each part of a program. However, this also requires the user to master the programming models of these architectures. Programming abstractions have been introduced to simplify the management of different types of devices, targeting both functional and performance portability.
Many approaches are devised as implementations of a heterogeneous task-based model. They present a common host-side API for orchestrating workloads/tasks, programmed as kernels, among the different accelerators present in the system.
Approaches such as Kokkos \cite{9485033}, OpenCL \cite{munshi2009opencl}, and OpenACC \cite{openacc} have become popular for mixing CPUs and GPUs. Other approaches also support the FPGAs with a similar high-level approach. 
%Approaches such as SYCL/oneAPI \cite{oneAPI}, OmpSs@FPGA \cite{bosch2018application}, and EngineCL \cite{nozal2020enginecl} have become popular for mixing CPUs, GPUs and FPGAs. 
However, despite improving general programmability by supporting a common host-side API, these approaches fail to provide the high flexibility potential of FPGAs. For example, these frameworks lack the support to independently swap in and out tasks of varying sizes onto an FPGA accelerator. The FPGA is programmed with a full bitstream that contains the kernels that will be run during the program execution in a non-preemptive way. % ARTURO: Avoid the term migration on this paper, it is a contribution for the journal paper
%and without the possibility of task migration. 
%Consequently, unlike the CPU or GPU, it is not possible to flexibly multiplex the FPGA between tasks, resulting in a more limited usage mode of the architecture and less opportunity for uptake in supercomputing.

The authors of~\cite{vaishnav2019heterogeneous}
explore these issues. They present a task-based model targeting System on-a Chip (SoC) deployment based on OpenCL and using DPR. They support kernel preemption by enabling checkpointing at the end of each OpenCL workgroup, and whilst this is a natural consistency point in the OpenCL model, the coarse-grained nature of the approach limits scheduling flexibility. For example, tasks of higher priority may need to wait until a previous workgroup with lower priority tasks finishes. Moreover, the user must write their kernel interfaces in a manner that are comformant to the interfaces of the Reconfigurable Regions (RR), % ARTURO ???of their design,
causing a conflict between the high-level OpenCL description and the management of the lower-level on-chip infrastructure, which increases the overall development complexity.
%and is ideally an aspect that would be abstracted.
%Furthermore, the use of the ZUCL framework \cite{pham2018zucl} to support their approach, which is a predefined static system, uses the entirety of the FPGA regardless of the number of kernels that are actually being executed by the application.

%Our approach unlocks flexibility in FPGAs by the use of a specific DPR capable design for each number of RRs, therefore avoiding potential decreases in the final design frequency due to opportunity missing in the routing phase. We provide a user-space C based driver for the PYNQ-Z2 FPGA that is leveraged by our heterogeneous abstraction Controller \cite{moreton2018controllers,rodriguez2021efficient}. This allows us to generate consistency points at arbitrary points during the execution of the kernel, instead of waiting for a specific consistency point, rendering our approach amenable for use cases based on priorities. Alongside the host-side abstraction we provide C99 macro based abstractions for writing the kernels that transparently perform code generation to adapt the user code to the interfaces of our RRs and to enable transparent preemption.

\section{The Controller programming model} \label{sect:controller}
Our proposal is devised as an extension of the Controller heterogeneous programming model. In this section we
provide and overview of the original model and its features. 
Controller \cite{moreton2018controllers,rodriguez2021efficient} is a heterogeneous task-based parallel programming model implemented as a C99 library. % ARTURO: Only part of the library are macro-functions
%based on preprocessor macros 
It provides an abstraction for programming using different types of devices, such as sets of CPU-cores, GPUs, and FPGAs.

\begin{wrapfigure}[14]{L}{0.7\textwidth}
\vspace{-28pt}
\begin{center}
    \centering
    \includegraphics[width=0.7\textwidth]{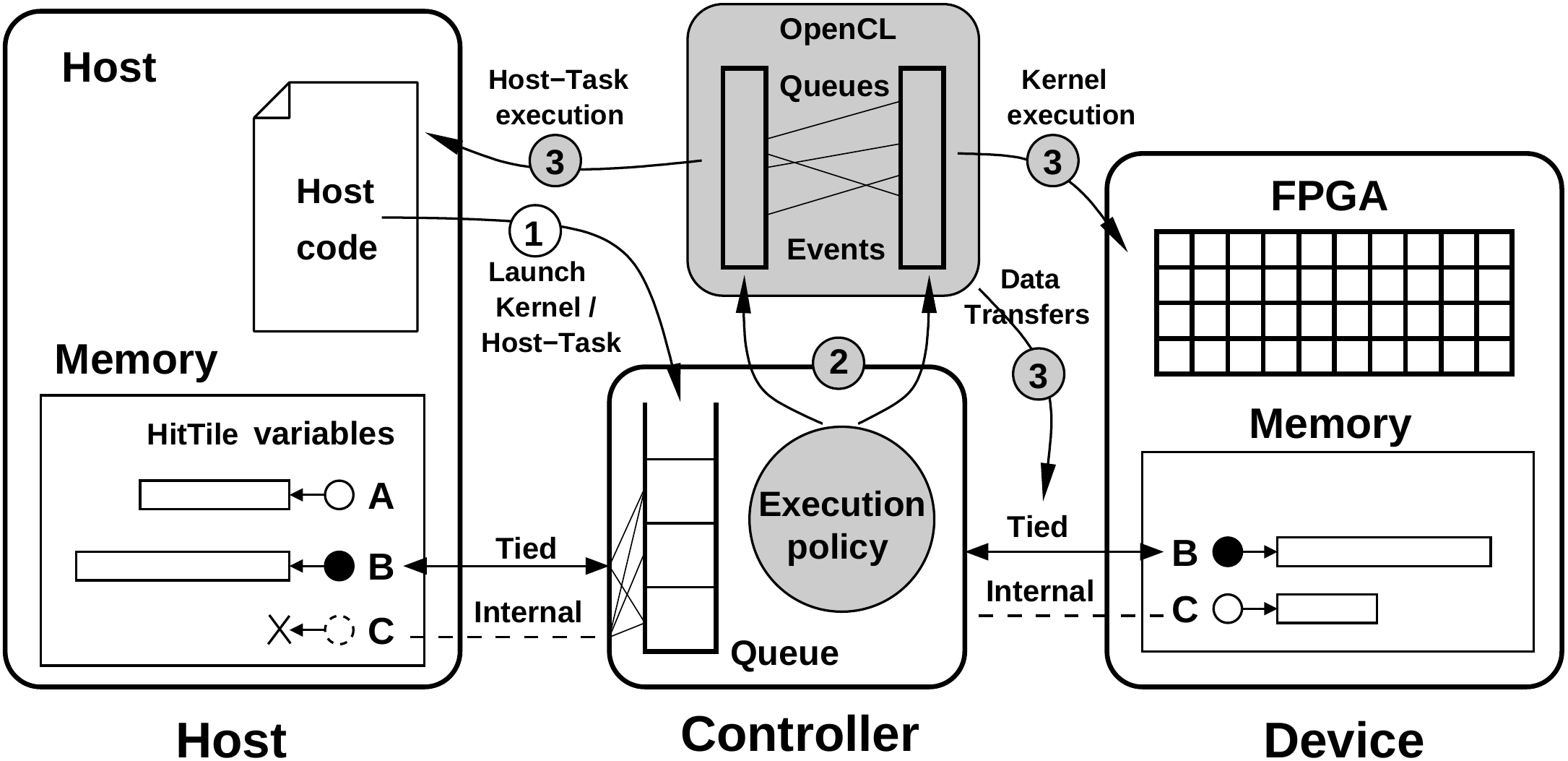}
    \caption{The Controller programming model, generic FPGA backend. Extracted from \cite{rodriguez2021efficient}.}
    \label{fig:controller}
\end{center}
\end{wrapfigure}

As illustrated in Figure~\ref{fig:controller}, the model is based around the Controller entity. Each Controller entity is associated to a particular device on its creation and manages the execution or data-transfers with that device.

The main program executes the coordination code in a main thread, using the Controller high-level API to enqueue computation tasks for the device. Each Controller entity has its own thread that dequeues and launches the execution of the kernels associated to the tasks. The model also provides an extra hidden device to execute host-tasks, which are managed through a separate thread. 
The Controller runtime resolves data dependencies between tasks automatically, performing data transfers in a transparent way.
The requests of both kernel executions and data-transfers needed are derived to the internal queues or streams of the device driver, controlling the execution order of kernels, data-transfers and host-tasks with native events. It uses three queues for each device: one for kernel execution, one for host-to-device transfer, and one for device-to-host transfers.
This enables a fast control operation and
an efficient overlapping of computation and data transfers when it is possible. Portability is achieved using different
runtime backends for different device technologies (such as CUDA or OpenCL), to implement the calls to manage the low-level device queues and events.
The computations that can be launched as tasks are kernel codes written by the programmer. Controller supports generic codes, written in OpenCL and targeting any kind of device, or specialised kernels programmed in the native programming model of an accelerator such as CUDA for Nvidia GPUs.

\section{Approach to support preemptive scheduling on FPGAs}
\label{sec:approach}
Our approach requires the use of new techniques in two areas, the on-chip FPGA infrastructure and the integration on the host-side of the Controller runtime.

%\subsection{Architecture of our approach} \label{sect:architecture}
\subsection{On-chip infrastructure} \label{sect:architecture}
\begin{wrapfigure}[16]{L}{0.7\textwidth}
\vspace{-10pt}
\begin{center}
    %\includesvg[width=\textwidth, inkscapelatex=false]{img/system_architecture}
    \includegraphics[width=0.7\textwidth]{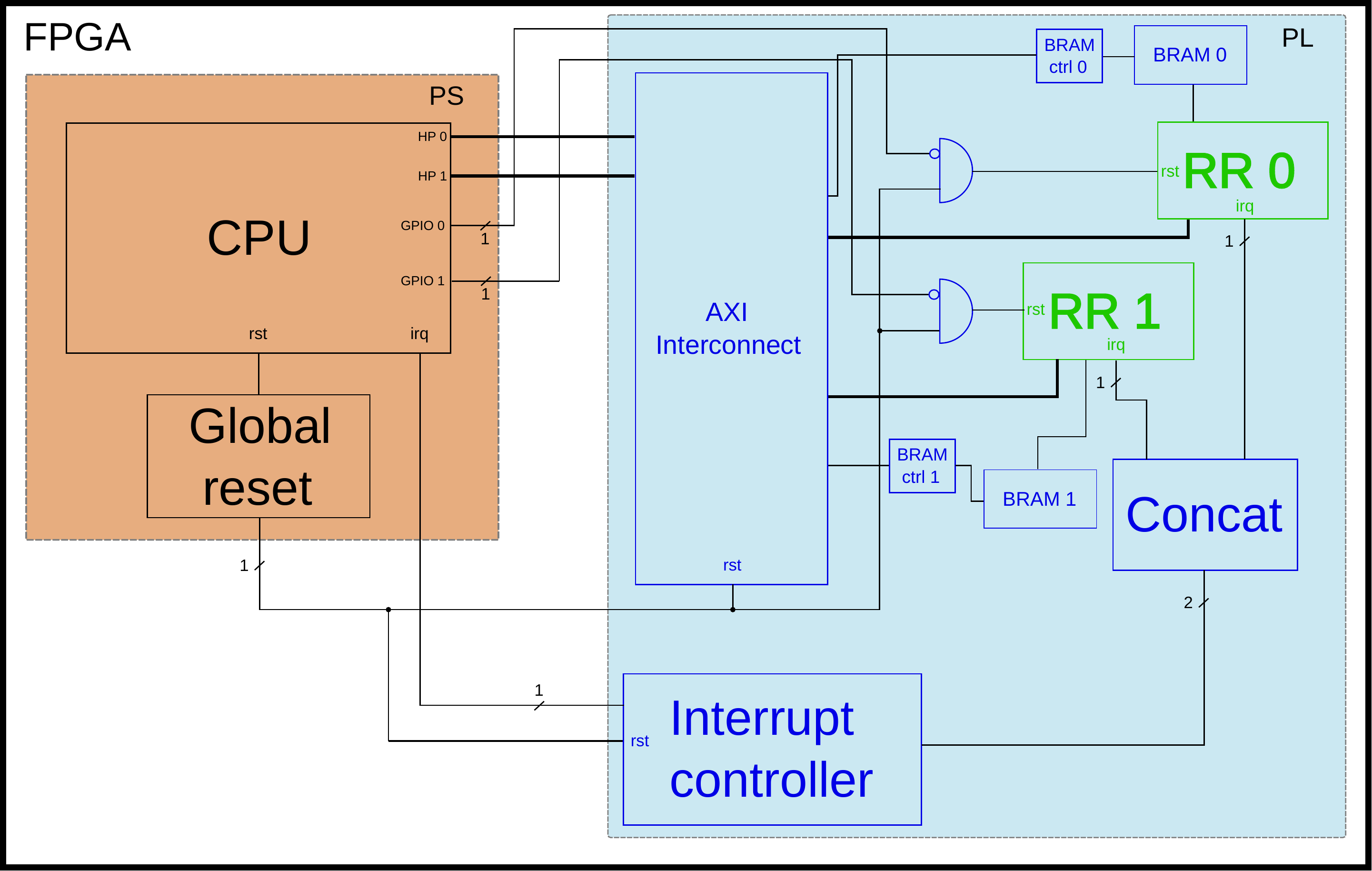}
    \caption{Simplified architecture of the system.}
    \label{fig:architecture}
\end{center}
\end{wrapfigure}

Figure~\ref{fig:architecture} shows the architecture of the static part of the on-chip infrastructure, known as \emph{shell},
that should be deployed in the FPGA to support the proposed control of Reconfigurable Regions (RR).
The example shows two RRs, although this model is scalable to any number of RRs.
The example shows details of a reference implementation of the proposal using Xilinx technology, although, these concepts can be easily ported to other FPGAs. This example shell implementation deploys HLS kernels generated by Xilinx's Vitis with 1 AXI4-Master interface which bundles the data ports to DRAM memory, and an AXI4-Slave interface bundling the control ports. %Any number of actual data ports can be used, as they are simply bundled into the appropriate data or control interface. 
This interface layout is fairly standard.
%The design illustrated in Figure \ref{fig:architecture} is known as the \emph{shell} and can be considered as %infrastructure. It is present on the FPGA throughout execution, with separate kernels, representing tasks, independently swapped into and out of the RRs. 
The interrupt controller registers interrupts generated by the RRs upon completion. Thus, the CPU can detect when kernels have finished execution. To support preemption, the shell should be able to interrupt a kernel, saving its context and state, to later resume it. The shell features two on-chip BRAM memory banks (one per RR) to store the interrupted kernels context at arbitrary intervals, defined by the user. BRAM memory is used since its speed and its closeness to the RRs results in very low latency, minimizing the overhead of the context saving operation. These BRAM banks are also connected to a BRAM controller which enables access from the CPU, supporting overall book-keeping of the kernel context when they are being swapped in and out by the scheduler that controls the execution from the host.

%Our approach increases complexity around resetting the state of the FPGA, where the ability to 
Our approach needs support for resetting both the entire FPGA and individual RRs (to undertake partial reconfigurations). The former is achieved via the shell's \emph{global reset} (see Figure~\ref{fig:architecture}). The latter is supported by a specific reset functionality for each RR.
%containing its own explicit reset functionality. 
%RR-specific resetting is required when undertaking a partial reconfiguration, and this is enabled by utilising the 
It is implemented using the GPIO ports of the CPU, with the added complexity that HLS kernels by default contain a low active reset. % meaning that 
We negate the GPIO signal and apply a logical \emph{and} with the global reset signal. The application of the reset signal is asynchronous which means that the kernel might be interrupted unpredictably. 
The software abstractions described in Section~\ref{sect:abst_preemption} ensure that the task can be resumed later 
from a consistent state.

%In order to undertake partial reconfiguration all the interfaces in the RR must be decoupled from the rest of the shell beforehand to prevent corruption of the logic or an undefined state \cite{dfx}. To achieve this all signals and buses connected to the RRs are filtered through a DFX Decoupler, which acts as a multiplexer setting the signals to a constant value in the interfaces when in decoupled mode and otherwise passes the data across unhindered.

This shell design is provided in \emph{netlist} form with the RRs instantiated as black boxes. Consequently, to generate the shell's bitstream the number of RRs required is supplied to the associated TCL script. This generates a corresponding Vivado compatible DPR capable hardware design, which is built and deployed onto the FPGA. The programmer writes their HLS kernels using our proposed software abstractions (see Section~\ref{sect:abst_preemption}), that effectively transform the C code into interface-compliant HLS code during the HLS synthesis. 

\subsection{Integration into the Controller framework}
A new backend has been written for the Controller framework. It supports interaction with our shell, targeting the Zynq-7020 FPGA in Pynq-Z2. %Whilst this is targeted at the Zynq-7020 present in the Pynq-Z2 used in our experiments of Section \ref{sec:experiments}, it is designed to be generic and support numerous FPGAs in the future. 
To communicate with the FPGA, our backend uses the Pynq C API \cite{pynq-c-api}. This API exposes low-level 
functionalities, such as the loading of both full and partial bitstreams, the interaction with design IP such as 
interrupt controllers or DMA engines through memory mapping, and host-device shared memory. 
Building on the C Pynq API means that this work is compatible with any other FPGA from the Zynq-7000 family with little modification required.

Each RR is treated as an independent accelerator by our backend to ensure that RR kernels can be executed in parallel. 
Thus, the Controller's queue is replicated as many times as the number of RRs, and each instance is managed by a separated thread. A request to reconfigure a region is implemented as an internal task, queued up and executed like any other task. This simplifies the backend structure and allows the scheduling of reconfigurations request before the associated task execution on the fabric. 
%This is slightly complicated by the fact that the 
Zynq only provides a single Internal Configuration Access Port (ICAP)~\cite{dfx}. This means that only one RR can be partially reconfigured at a time. Thus, we need to implement
%resulting in the requirement for 
a synchronisation between reconfiguration request in the Controller queues.
The Zynq-7000 FPGA family architecture supports shared memory which can be accessed by both the FPGA fabric and host CPU.
Thus, data-movement operations can be implemented with zero-copy. 
The backend utilises Userspace I/O (UIO) to interact with the shell's interrupt controller to detect the interrupts 
%when these are
raised by the RRs to indicate kernel termination.
%Unlike a polling approach, which would keep the CPU's thread running on a core, instead 
We use the \emph{select()} system call to activate the manager CPU thread when an interrupt is received.
Then, the backend queries the interrupt controller to determine which RR raised the interrupt.
This avoids the use
of an active polling approach that would keep a CPU core busy unnecessarily.

\subsection{Use case: DPR scheduler} \label{sect:dpr_scheduler}
In this section we show the use the proposed DPR approach to build an FCFS scheduler of kernel tasks, with priorities and preemption. 

In this proof-of-concept we simulate scenarios where both the time of the next task arrival and the task parameters are randomly generated. We pre-generate a sequence of tasks (\texttt{tasks\_to\_arrive}), ordered by a random arrival time.
Each task has a random priority, a randomly chosen kernel code to execute (from a given set), and random arguments.
We design a modular scheduler with separate modules for the generation of random tasks, management of the queues, service of tasks and the main loop of the scheduler. Therefore, it is easy to extend or adapt. 
It is compatible with any number of RRs.
%and the underlying configuration for management of partial reconfiguration is generated transparently from the kernels specified by the user.
%, since the name of each kernel can be extracted from \texttt{CTRL\_KERNEL\_FUNCTION}.

\begin{wrapfigure}[14]{L}{0.65\textwidth}
\vspace{-45pt}
\begin{minipage}{0.65\textwidth}
\begin{algorithm}[H]
    \caption{Main loop of the scheduler.}
    \label{alg:man_loop_scheduler}
    \begin{algorithmic}
        \While{$true$}
            \State {$Ctrl\_WaitForInterrupt(\&timeout)$}
            \If {$has\_finished(N, R, \&tasks\_to\_arrive$)}
                \State {break}
            \EndIf
            
            \If {$tasks\_to\_arrive \; \&\& \; timeout == 0$}
                \State $task = get\_arrived\_task()$
            \Else
                \State $task = get\_task\_from\_queue()$
            \EndIf
            \State $serve\_task(task, R, P)$
            \State $update\_timeout(\&timeout)$
        \EndWhile
    \end{algorithmic}
\end{algorithm}
\end{minipage}
\end{wrapfigure}

The main loop of the scheduler is presented in Algorithm \ref{alg:man_loop_scheduler}. The arrival of the next task 
is simulated with a timeout clock, used in the same \texttt{select()} function that detects the interrupts raised
by the end of a kernel in a RR. Thus, the \texttt{WaitForInterrupt} function returns when a new task arrives or when a RR kernel finishes.
%When a new task arrives, the module to generate it is called.
%In every iteration a task is served.
%, either just arrived/generated or retrieved from a queue, is served.

The process of serving a task consists of the following steps:
%\begin{enumerate}
%    \item 
    (1) Find an available region, i.e., a region where the last task running has already finished. \label{step:sch_available}
    (2) In case no available region was found, if preemption is disabled enqueue the task.
    %go to step \ref{step:sch_enqueue}. 
    If preemption is enabled, check if there is a region executing a task with lower priority.
    %, a task can be preempted from a region where its running task has a lower priority than the incoming task. This region is now considered available. Go to step \ref{step:sch_available}.
    In that case, stop the kernel execution in that region, save the context and state, enqueue the stopped
    task, and consider the
    region as available.
    (3) If the kernel loaded in the available region is distinct from the kernel of the incoming task, enqueue a swapping task to reconfigure the RR. \label{step:sch_swap}
    (4) Launch the new task. If it was a previously stopped task, its context is copied back to the device before
            launching.
  % \item Enqueue the incoming task and finish. \label{step:sch_enqueue}
%\end{enumerate}

%Finally, 
%The $update\_timeout()$ function loads the arrival time of the next task in the $timeout$ structure when the last incoming task has been served.

%The scheduler is modular, with separate modules for the generation of tasks, management of the queues, service of tasks and the main loop of the scheduler, therefore easy to extend. It is compatible with any number of RRs and the underlying configuration for management of partial reconfiguration is generated transparently from the kernels specified by the user, since the name of each kernel can be extracted from \texttt{CTRL\_KERNEL\_FUNCTION}.

\section{Programmer's abstractions}
\label{sec:abstractions}
This section describes the abstractions provided to the programmer to implement kernels and to use the
proposed approach, without knowledge of the low-level details of the DPR technology.

\subsection{Kernel interface abstraction}
The generation of interfaces in technologies such as Vitis HLS is done adding pragmas that can be cumbersome and error prone to write. Moreover, a requirement of DPR is that HLS kernels to be deployed into a given RR must present the same external interface to the shell. They must conform to the same number of interface ports and port configurations, such as bus widths~\cite{dfx}. Thus, better abstractions are needed to hide these low-level details to the programmer.

\begin{wrapfigure}[19]{L}{0.55\textwidth}
\vspace{-20pt}
\begin{lstlisting}[language=C, basicstyle=\scriptsize, caption={Sketch of a Median Blur kernel written with the Controller abstraction}, label={lst:median_blur}]
@\textcolor{green}{CTRL\_KERNEL\_FUNCTION(
  MedianBlur, PYNQ, DEFAULT, 
  KTILE\_ARGS(KHitTile\_int input\_array,
             KHitTile\_int output\_array), 
  INT\_ARGS(int H, int W, int iters), 
  FLOAT\_ARGS(NO\_FLOAT\_ARG)) }@ {
  ...
  int k, row, col;
  @\textcolor{green}{context\_vars(k, row, col);}@
  ...
  @\textcolor{green}{for\_save(k, 0, iters, 1)}@ { @\label{code:for_save1}@
    @\textcolor{green}{for\_save(row, 1, H+1, 1)}@ { @\label{code:for_save2}@
      @\textcolor{green}{for\_save(col, 1, W+1, 1)}@ { @\label{code:for_save3}@
        window[0] = hit(
    input_array, row-1, H_NCOL+col-1);
        ...
        @\textcolor{green}{checkpoint(col);}@ @\label{code:checkpoint1}@
      } @\textcolor{green}{checkpoint(row);}@ @\label{code:checkpoint2}@
    } @\textcolor{green}{checkpoint(k);}@ @\label{code:checkpoint3}@
  }   
}
\end{lstlisting}
\end{wrapfigure}

The configuration of the interfaces is a parameter present in our TCL configuration script that generates the shell's hardware design, as discussed in Section \ref{sect:architecture}. % In the experiments reported in this paper we have configured our shell such that the RRs contain one AXI-Master interface and one AXI-slave, providing a single data and control port. %As discussed in Section [REF], one can bundle multiple arrays into a single port supporting an arbitrary number of inputs and outputs to the kernel, albeit this bundling potentially resulting in some contention depending on the access pattern. 
In the Controller model, the kernel codes are wrapped with curly brackets and preceded by a kernel signature.
The kernel signature is provided with a macro-function named
\texttt{CTRL\_KERNEL\_FUNCTION}.
It specifies the kernels parameters in a form that is processed by the Controller library to generate
the proper low-level interface.
Listing~\ref{lst:median_blur} illustrates the definition of a Median Blur kernel, used in our evaluation in Section \ref{sec:experiments}, preceded by its signature.
In this work we extend the Controller kernel signature to generate code with a uniform interface, as required by the
shell. The parameters of the kernel signature are the following:

%\texttt{CTRL\_KERNEL\_FUNCTION(K, T, S, A_{p}, A_{i}, A_{f})}: 
\begin{itemize}
\item[] \texttt{CTRL\_KERNEL\_FUNCTION(K, T, S, A\textsubscript{p}, A\textsubscript{i}, A\textsubscript{f})}: 
\begin{itemize}
     \item \texttt{K} is the name of the kernel.
     \item \texttt{T} indicates the backend type that will be targeted. Supported types are: CPU, CUDA, OpenCL, FPGA.
     \item \texttt{S} is the subtype of backend that will be targeted, e.g. \texttt{DEFAULT}.
     \item \texttt{A\textsubscript{p}} is a list of pointer non-scalar arguments defined with \texttt{KTILE\_ARGS}.
     \item \texttt{A\textsubscript{i}} is a list of integer scalar arguments defined with \texttt{INT\_ARGS}.
     \item \texttt{A\textsubscript{f}} is a list of floating point scalar arguments defined with \texttt{FLOAT\_ARGS}.
 \end{itemize}
\end{itemize}

Controller provides a wrapper structure for multi-dimensional arrays named \emph{HitTile}. 
Any kind of non-scalar arguments
are provided as \emph{HitTile} arguments. \texttt{KTILE\_ARGS} function enables the use of \emph{HitTile} accessors within the kernel, effectively providing input and output arrays to the kernel, as discussed in \cite{rodriguez2021efficient}. \texttt{INT\_ARGS} and \texttt{FLOAT\_ARGS} support passing integer and float scalar arguments, respectively. All these
functions have variadic arguments to adapt the kernel interface to the number of arguments required by the programmer.
%\texttt{KTILE\_ARGS}, \texttt{INT\_ARGS} and \texttt{FLOAT\_ARGS} argument macros are variadic, which enables them to adapt to the number of arguments provided by the user. Listing \ref{lst:median_blur} illustrates the definition of the Median Blur kernel, used in our evaluation of Section \ref{sec:experiments}, where this use of the \texttt{CTRL\_KERNEL\_FUNCTION} macro to define the interface can be seen. 
The corresponding code generated by the kernel signature for the kernel shown in Listing~\ref{lst:median_blur} is shown
in Listing \ref{lst:medianblur_signature}. Three integer arguments are provided by the user and five extra dummy
arguments \texttt{i\_args\_<n>} are generated. Similarly 8 dummy floating point and 1 dummy pointer arguments are generated to fill the argument count and provide a shell compliant interface. Finally, a pointer to a \texttt{struct context} is added for context book-keeping if the task is interrupted. %and \texttt{int * return\_var} for the simulation of a return variable are%
%is generated. %This allows returning values integer values in the Controller model, that requires kernel to be \texttt{void} functions. This can be easily extended for other returning other types by extending the signature macro.

% \begin{lstlisting}[language=C, basicstyle=\scriptsize, label={lst:medianblur_signature}, caption={Code generation for the signature of the Median Blur kernel}]
% void MedianBlur(
%     wrapper_int input_array_wrapper, data_int input_array_data, 
%     wrapper_int output_array_wrapper, data_int output_array_data, 
%     wrapper_int ktile_args_0_wrapper, data_int ktile_args_0_data, 
%     int H, int W, int iters, int i_args_0, int i_args_1, int i_args_2, int i_args_3, int i_args_4, 
%     float no_float_arg, float f_args_0, float f_args_1, float f_args_2, float f_args_3, float f_args_4, float f_args_5, float f_args_6, 
%     volatile struct context * context, int * return_var);
% \end{lstlisting}
\begin{lstlisting}[language=C, basicstyle=\scriptsize, label={lst:medianblur_signature}, caption={Code generation for the signature of the Median Blur kernel}]
void MedianBlur(..., 
    int H, int W, int iters, int i_args_0, ..., int i_args_4, 
    ..., volatile struct context * context, int * return_var);
\end{lstlisting}

%The generation of interfaces in Vitis HLS happens through the use of pragmas that can be cumbersome and error prone to write. In Controller these are abstracted through the interface definition macros \texttt{DEF\_<type>\_INTERFACES}, where $type$ is one of \texttt{KTILE} (for pointer interfaces), \texttt{INT}, \texttt{FLOAT} or \texttt{RETURN}. The user provides the list of arguments for which the interface pragmas are to be generated or the keyword \texttt{NO\_<type>\_INTERFACE} if there is no argument of that type in the kernel.

%%%%
\subsection{Programmer abstractions for preemption} \label{sect:abst_preemption}
% data is easy - scalars not so! (internal anyway)

Preemption of a kernel whilst it is running requires saving its state so that it can be resumed in the future.
Previous approaches, such as~\cite{vaishnav2019heterogeneous} only save the context at the end of an OpenCL workgroup.
We also wanted to provide flexibility for the programmer to decide exactly where their code should be checkpointed. 
We propose a finer-grain and programmer-aware checkpointing approach, where the programmer has the flexibility to indicate when and what data should be chekpointed during the kernel execution.
%The kernel's external data, for instance input and output arrays, is fairly easy to handle as these are stored in explicit memory areas that can be accessed by both the FPGA and CPU, and hence conveniently accessed by both to obtain the current state of the kernel. However internal data, such as scalars, are much more difficult to handle as these are often not represented by the HLS tooling as memory, as one would expect with Von Neumann architectures, but instead might be a wire on the dataflow graph connecting one construct with another. Consequently checkpointing these is significantly more complex, with some approaches explicitly modifying the generated hardware description language to support such activities [REF]. 
%It was our objective to support preemption at the HLS level, without any changes required to the generated code. 
%Furthermore, unlike 
%Previous approaches, such as~\cite{vaishnav2019heterogeneous} only save the context at the end of an OpenCL workgroup.
%we also wanted to provide flexibility for the programmer to decide exactly where their code should be checkpointed. 
We provide several checkpointing macro-functions. 
%(see Listing \ref{lst:median_blur}).
The programmer declares which variables should be stored in the checkpoints using the \texttt{context\_vars} macro-function. The \texttt{checkpoint} macro stores one or more of these variables at a given execution point. A \texttt{for\_save} macro-function is used in-place of the normal \emph{for} loop construct, to provide support for resumption on a specific loop iteration.
These calls are expanded to the proper code at synthesis time. 
%to store the values on fast on-chip BRAM memory.

%, and the programmer 
%Listing \ref{lst:median_blur} sketches the Median Blur kernel code used for our experiments in Section \ref{sec:experiments} which provides an illustration of these checkpoint-restart macros in use.
An example of their use is shown in Listing \ref{lst:median_blur}.
At line 11 the integer variables \emph{k}, \emph{row}, and \emph{col} are selected to be checkpointed, with lines \ref{code:for_save1}, \ref{code:for_save2}, and \ref{code:for_save3} using the \texttt{for\_save} macro to define loops and for these to be restarted as appropriate. The associated loop variables are checkpointed at lines \ref{code:checkpoint1}, \ref{code:checkpoint2}, and \ref{code:checkpoint3}. 
This kernel saves the state a each iteration to be able to be resumed without discarding previously computed
iterations.
%In this example we have chosen to save this state at every iteration, this means that we are able to resume execution without discarding any of the previously computed iterations, although this might result in some storage overhead as described in Section \ref{sect:architecture} this is minimised by our shell using the fast on-chip BRAM memory for this checkpointing.

\begin{wrapfigure}[9]{l}{0.27\textwidth}
\vspace{-15pt}
\begin{lstlisting}[language=C, basicstyle=\scriptsize, label={lst:struct_context}, caption={Definition of \texttt{struct context}}.]
struct context {
  int var[N];
  int init_var[N];
  int incr_var[N];
  int saved[N];
  int valid;
};
\end{lstlisting}
\end{wrapfigure}

Context saving is done 
transparently storing the state 
%at intervals indicated by the macros 
in the \texttt{struct context} generated in BRAM (see Listing~\ref{lst:struct_context}. 
In our prototype up to \emph{N} integers can be nominated by the user to be saved, where $N$ is a compile time parameter. It is trivial to extend the structure to support other data types. 
The field \texttt{saved} keeps information about 
%informs the \texttt{context\_vars} macro 
whether the variables have already been saved through checkpoint and they should be restored in a resume operation.
%and if their last value before preemption should be used. 
%If the value is 0 
%then this is the first execution of the kernel and the checkpointed 
%the variables should be set to the programmer's initial values specified in the code.
%As discussed in Section \ref{sect:architecture}, since preemption is asynchronous 
The \texttt{valid} field is used to indicate if the asynchronous preemption interrupted the kernel during a data
saving operation. In that case, the resume operation will be done with the previously saved values.
%used as a lock between code sections in which the context structure is being modified. If the field is 1 after reset, then the information in the structure is valid, otherwise it is corrupted and the last iteration will be discarded when the kernel is resumed by manipulating the values of the loop variables in the structure.

\section{Experimental study}
\label{sec:experiments}
We present the results of an experimental study to evaluate the efficiency of our approach.

\subsection{Use case: Scheduler of randomly generated image filter tasks}
In this study we experiment with
%The work presented in this paper is assessed through 
the scheduler described in Section \ref{sect:dpr_scheduler}. 
The kernels chosen for the experimentation are blur image filters applied to images pre-stored in memory. 
%this experiment were the Median Blur and Gaussian Blur filters. 
Tasks execute one of four possible kernels: Median Blur over, two or three iterations, or one iteration of Gaussian Blur. %The filters are applied over randomly generated tasks arriving at 
Tasks arrive at random times distributed over $\unif(0,T)$ minutes.
The scheduler features optional preemption and priorities.
For these experiments we choose to use 5 different priorities, to generate enough preemptions, task switching and reconfigurations.
%we use and in this section we opt for five as this setting is high enough to study the effects of the scheduling on the queues.
%The tasks associated with each of our four kernels, along with their arrival time, and input and output images are randomly generated before the scheduler starts. 
%The tasks associated with each of our four kernels, along with their arrival time and input and output images, are randomly generated before the scheduler starts. 
The tasks, their arrival time, and the image on which it should be applied, are randomly generated before the scheduler starts.

\subsection{Experimentation environment}
The experiments were conducted on a Xilinx PYNQ-Z2 FPGA. It features a ZYNQ XC7Z020-1CLG400C of the Zynq-7020 family, an ARM Cortex-A9 dual core at 650 MHz CPU and 512 MB DDR3. HLS kernels were compiled using Xilinx Vitis HLS version 2020.2 and the hardware design and corresponding bitstreams were generated with Xilinx Vivado v2020.2. Controller was compiled with GCC 9.3.0 and compilation scripts were generated with CMake 3.20.5.

Several random seeds for the task generation have been tested. We show the results for the value 15. The main observations can be extrapolated for other random sequences.
%for the results for service time, and this is defined as the amount of time taken for a task to start execution on a RR after its arrival. This is useful as it illustrates an important scenario in which preemption is beneficial due to the priorities of the incoming tasks. Although not shown here the scheduler was tested with a range of seeds to ensure generality. 
The number of tasks generated was chosen to be 30.
We enabled priorities both with and without preemption of tasks. We considered three different rate of arrivals $T$: busy (0.1), medium (0.5) and idle (0.8). We worked with image sizes $200\times200$, $300\times300$, $400\times400$, $500\times500$ and $600\times600$. In order to study the sequential vs. the parallel behaviour both one and two RRs were considered. Finally, each experiment was executed ten times to account for variability and the results presented are average times with standard deviation.

\subsection{Results}
\begin{figure}
    \centering
    \begin{subfigure}{0.48\textwidth}
    \includegraphics[width=\textwidth]{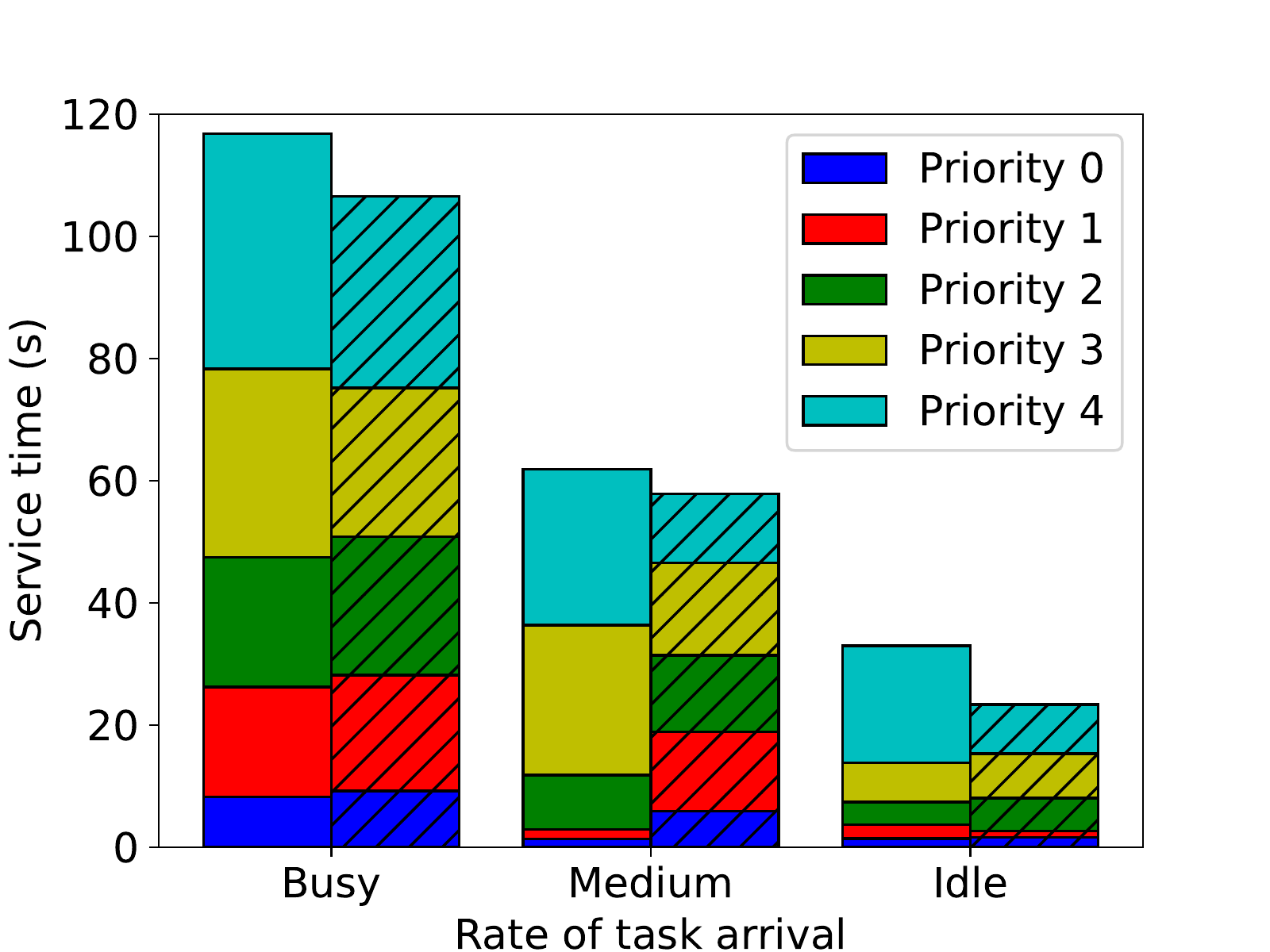}
    \label{fig:service_time_1rr}
    \end{subfigure}
    \begin{subfigure}{0.48\textwidth}
    \includegraphics[width=\textwidth]{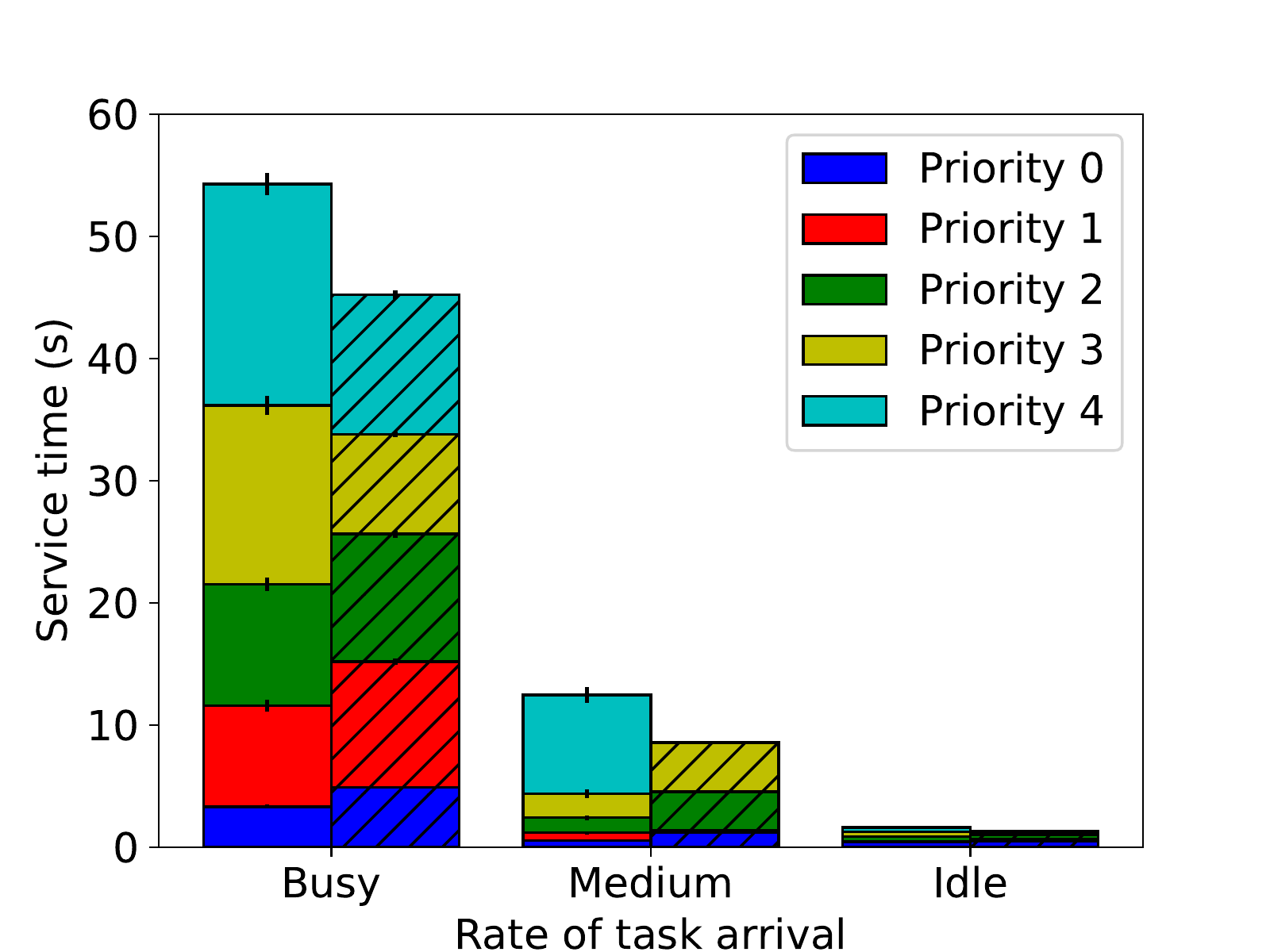}
    \label{fig:service_time_2rr}
    \end{subfigure}
    \caption{Service times for 30 tasks at size $600 \times 600$. 1 RR (left), 2 RRs (right). Per bar group: Non-preemptive (left), preemptive (right).}
    \label{fig:service_times}
\end{figure}

In order to show the effectiveness of our approach we are presenting results for the following metrics: (i) service time, defined as the time it takes for a task to be served since it is generated until it starts execution on the FPGA and (ii) throughput, defined as the number of tasks executed per second. %Note that (i) only measures the time for the first service and does not take into account the time for further services after preemption, since this metric intends to measure the effectiveness of the scheduler to serve urgent tasks. 
We also compare the use of partial reconfiguration with the more conventional full reconfiguration approach.
Figure \ref{fig:service_times} reports the service time for tasks in every priority queue both with and without preemption for 30 tasks at size $600\times600$ accumulated by priority. We chose this number of tasks and image size as it provides enough workload and a sufficient number of tasks to study the behaviour of the scheduler. The results are presented both for one and two RRs. As can be seen service times are longer for the busy rate of arrival than for medium and idle, as tasks have to wait a longer time until a RR becomes available than when they arrive later, giving the opportunity for kernels to finish. If the priority of an incoming task is higher than one of the tasks running, then its service time will be virtually zero. We can observe this by comparing the plots on the right with plots on the left. For this representative case, on average, preemption reduces service time substantially. This will be the case in general when incoming tasks present a higher priority than running tasks. These results show that our scheduler effectively reduces the total service time of tasks, thus increasing the flexibility, as preemption enables swapping in and out tasks upon a condition --- priority in this case. The reduction in service time is heavily dependent on the structure of priorities of the generated tasks, both in terms of the number of tasks enqueued and the number of reconfigurations enforced by incoming kernels not loaded already in the fabric. 
Note that a task will have to wait until previous tasks of higher or the same priority have completed.
%This can create the effect that even highest priority tasks may starve, as other highest priority tasks might have arrived before. 
Additionally, as shown in Figure \ref{fig:service_times}, the service time decreases with the number of RRs, as more opportunities are created for kernels of lower priorities to execute.
%that can be evicted.
%and available RRs are created.

% \begin{figure}[thb]
% \begin{subfigure}{0.5\textwidth}
%     %\centering
%     \includegraphics[width=\textwidth]{plots/service_times-no-preemption-1rr-600-15.pdf}
%     \caption{1 RR without preemption.}
%     \label{fig:my_label}
% \end{subfigure}
% \begin{subfigure}{0.5\textwidth}
%     %\centering
%     \includegraphics[width=\textwidth]{plots/service_times-preemption-1rr-600-15.pdf}
%     \caption{1 RR with preemption.}
%     \label{fig:my_label}
% \end{subfigure}
% % \caption{Service times for 30 tasks launched on 1 RR.}
% % \end{figure}

% % \begin{figure}
% \begin{subfigure}{0.5\textwidth}
%     %\centering
%     \includegraphics[width=\textwidth]{plots/service_times-no-preemption-2rr-600-15.pdf}
%     \caption{2 RR without preemption.}
%     \label{fig:my_label}
% \end{subfigure}
% \begin{subfigure}{0.5\textwidth}
%     %\centering
%     \includegraphics[width=\textwidth]{plots/service_times-preemption-2rr-600-15.pdf}
%     \caption{2 RR with preemption.}
%     \label{fig:my_label}
% \end{subfigure}
% \caption{Service times for 30 tasks at size $600\times600$.}
% \label{fig:service_times}
% \end{figure}

\begin{figure}
    \centering
    \includegraphics[width=0.8\textwidth]{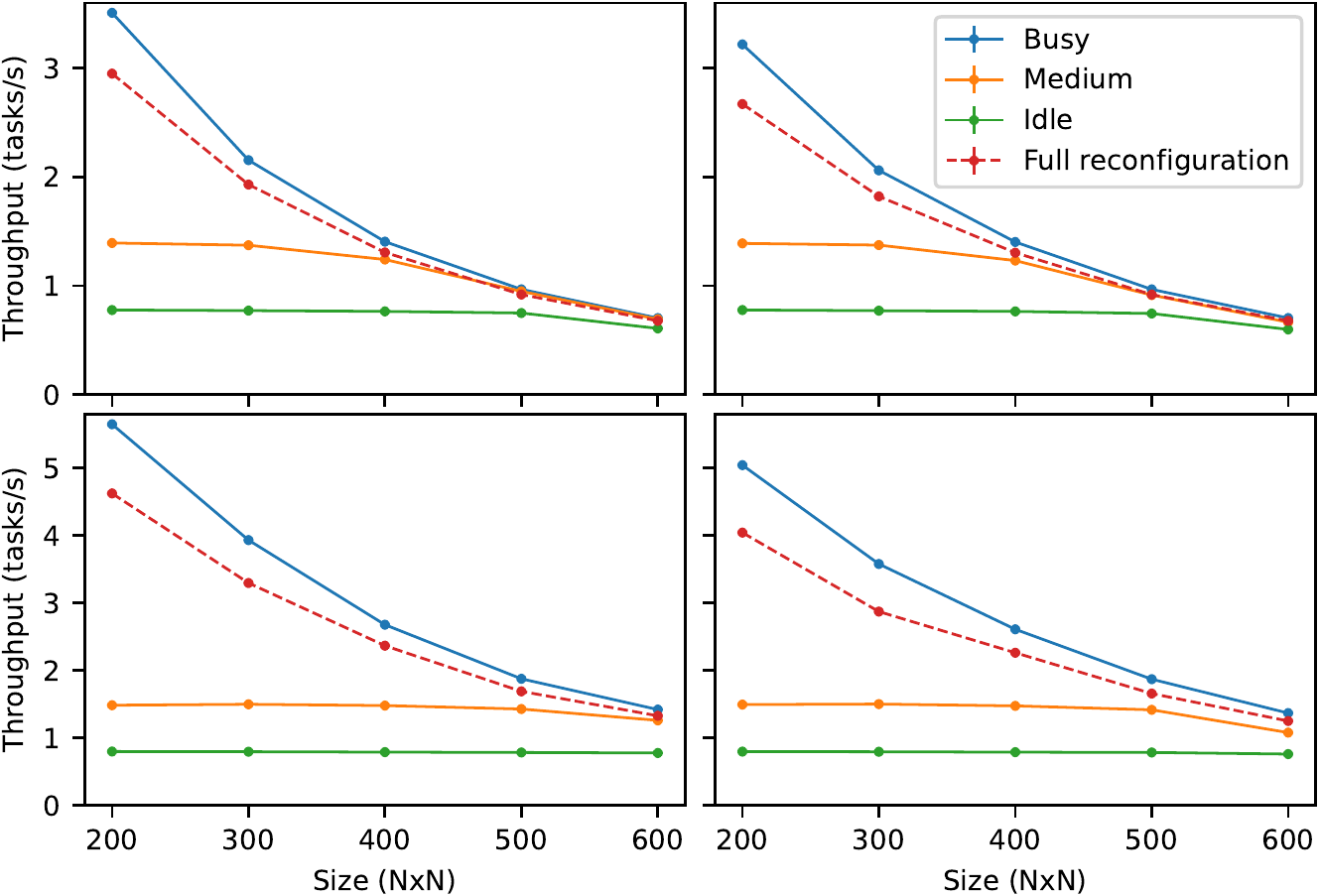}
    \caption{Throughput for 30 tasks. 1 RR (first row) and 2 RRs (second row). Non-preemptive (first column) and preemptive (second column).}
    \label{fig:throughput}
\end{figure}

Figure \ref{fig:throughput} shows the throughput of the scheduler with 30 tasks both with and without preemption over one and two RRs. As expected, the throughput increases with the rate of arrival of tasks. The lower the dimensions of the images the higher the throughput, as the kernels complete execution faster. It is also noticeable that the overheads incurred by preemption lead to a slightly lower throughput. These are most noticeable for a high rate of arrival of tasks, where throughput losses of 8.3\% and 10.7\% for the case with one and two RRs, respectively, at size 200 and busy arrival rate.
%are observed.
For the rest of cases the loss ranges between 0-4\%. Most of this overhead is explained by the time taken by the extra partial reconfigurations imposed by preemption. The dashed red lines show an upper bound of the throughput if full reconfiguration was used instead. This has been calculated from the throughput at busy rate of arrival adding the product of the number of reconfigurations by the average difference on time between full (0.22 s) and partial (0.07 s) reconfiguration. This is a highly optimistic upper bound, since it does not take into account the effects of stalling the FPGA, which impedes the concurrency of kernel execution and reconfiguration, and enforces a preemption of the rest of kernels that are to be kept in the FPGA. Finally, the average preemption overhead observed is 1.66\% for one RR with standard deviation 2.60\%, and 4.04\% for two RRs with a standard deviation of 7.16\%. The deviation is high due to a overhead peak of 23.40\% for busy rate of arrival at size $200\times200$. This indicates that this technique might not be interesting for short tasks whose execution time is comparable to the reconfiguration time. 

% \begin{figure}[htb]
% \begin{subfigure}{0.5\textwidth}
%     %\centering
%     \includegraphics[width=\textwidth]{plots/throughput-1rr-no-preemption.pdf}
%     %\caption{1 RR without preemption.}
%     \label{fig:my_label}
% \end{subfigure}
% \begin{subfigure}{0.5\textwidth}
%     %\centering
%     \includegraphics[width=\textwidth]{plots/throughput-1rr-preemption.pdf}
%     %\caption{1 RR with preemption.}
%     \label{fig:my_label}
% \end{subfigure}

% \begin{subfigure}{0.5\textwidth}
%     %\centering
%     \includegraphics[width=\textwidth]{plots/throughput-2rr-no-preemption.pdf}
%     %\caption{2 RR without preemption.}
%     \label{fig:my_label}
% \end{subfigure}
% \begin{subfigure}{0.5\textwidth}
%     %\centering
%     \includegraphics[width=\textwidth]{plots/throughput-2rr-preemption.pdf}
%     %\caption{2 RR with preemption.}
%     \label{fig:my_label}
% \end{subfigure}
% \caption{Throughput for 30 tasks.}
% \label{fig:throughput}
% \end{figure}

\section{Conclusions} \label{sect:conclusions}
This work presents a task-based abstraction for programming FPGAs that enables task preemption using DPR. We abstract the low-level details of the generation of a DPR capable system and provide a high-level API for simple management of kernel launch, data transfer and transparent book-keeping for context preemption. We show that our approach enhances flexibility by reducing the service time of urgent tasks thanks to the ability to swap tasks in and out. The overhead of preemptive vs. non-preemptive scheduling with DPR is 1.66\% on average for one RR and 4.04\% for two RRs. Finally, our simulations show significant performance gains over the traditional use of full reconfiguration.

Future work includes, in no particular order:
\begin{enumerate}
    \item Task migration between FPGA and other architectures e.g. GPU and CPU.
    \item Extension to data-center FPGAs e.g. as Xilinx Versal and Xilinx Alveo.
    \item Extension of the backend to leverage full reconfiguration to provide an accurate measure of the performance gain through the use of DPR.
    \item Reduction of the overhead of this technique with a custom ICAP controller, as Xilinx's can only exploit up to 2.5\% of the port bandwidth \cite{vipin2018fpga}. 
\end{enumerate}

\bibliographystyle{plain}
% \nocite{*}
\bibliography{bibliography/bibliography}

\end{document}